\documentclass[10pt,a4paper,twocolumn,american,superscriptaddress,aps,prd,nofootinbib]{revtex4-1}
\usepackage{lmodern}

\usepackage[T1]{fontenc}
\usepackage[utf8]{inputenc}
\usepackage{color}
\usepackage{babel}
\usepackage{amsmath}
\usepackage{amssymb}
\usepackage{esint}
\usepackage[unicode=true,
 bookmarks=true,bookmarksnumbered=false,bookmarksopen=false,
 breaklinks=false,pdfborder={0 0 1},backref=false,colorlinks=true]
 {hyperref}
\hypersetup{pdftitle={Observables and unobservables in dark energy cosmologies},
 pdfauthor={Luca Amendola, Martin Kunz, Mariele Motta, Ippocratis D. Saltas, Ignacy Sawicki},
 pdfkeywords={dark energy, modified gravity, large-scale structure},
 citecolor=blue,filecolor=blue,linkcolor=blue,urlcolor=blue}

\makeatletter

\pdfpageheight\paperheight
\pdfpagewidth\paperwidth

 \@ifundefined{textcolor}{}
 {%
   \definecolor{BLACK}{gray}{0}
   \definecolor{WHITE}{gray}{1}
   \definecolor{RED}{rgb}{1,0,0}
   \definecolor{GREEN}{rgb}{0,1,0}
   \definecolor{BLUE}{rgb}{0,0,1}
   \definecolor{CYAN}{cmyk}{1,0,0,0}
   \definecolor{MAGENTA}{cmyk}{0,1,0,0}
   \definecolor{YELLOW}{cmyk}{0,0,1,0}
 }
\usepackage{enumitem}		% customizable list environments
\makeatother

\begin{document}

\title{Observables and unobservables in dark energy cosmologies}

\author{Luca Amendola}

\affiliation{ITP, Ruprecht-Karls-Universität Heidelberg, Philosophenweg 16, 69120
Heidelberg, Germany}

\author{Martin Kunz}

\affiliation{Département de Physique Théorique and Center for Astroparticle Physics,
Université de Genève, Quai E.\ Ansermet 24, CH-1211 Genève 4, Switzerland}

\author{Mariele Motta}

\affiliation{ITP, Ruprecht-Karls-Universität Heidelberg, Philosophenweg 16, 69120
Heidelberg, Germany}

\affiliation{Instituto de Física Gleb Wataghin -- UNICAMP, 13083-970 Campinas,
SP, Brazil}

\author{Ippocratis D. Saltas}

\affiliation{School of Physics \& Astronomy, University of Nottingham, Nottingham,
NG7 2RD, United Kingdom}

\author{Ignacy Sawicki}

\affiliation{ITP, Ruprecht-Karls-Universität Heidelberg, Philosophenweg 16, 69120
Heidelberg, Germany}

\begin{abstract}
The aim of this paper is to answer the following two questions: (1)
Given cosmological observations of the expansion history and linear
perturbations in a range of redshifts and scales as precise as is
required, which of the properties of dark energy could actually be
reconstructed without imposing any parameterization? (2) Are these
observables sufficient to rule out not just a particular dark energy
model, but the entire general class of viable models comprising a
single scalar field?

This paper bears both good and bad news. On one hand, we find that
the goal of reconstructing dark energy models is fundamentally limited
by the unobservability of the present values of the matter density
$\Omega_{\text{m}0}$, the perturbation normalization $\sigma_{8}$
as well as the present matter power spectrum. On the other, we find
that, under certain conditions, cosmological observations can nonetheless
rule out the entire class of the most general single scalar-field
models, i.e.~those based on the Horndeski Lagrangian. 
\end{abstract}

\date{\today}

\maketitle

\section{Introduction}

Research in dark energy (DE) cosmology is generally devoted to building
viable models and to constraining them from observations (see, for
instance, the reviews \cite{Amendola2010,Clifton:2011jh,Kunz:2012aw}).
The models are usually characterized by a small number of properties
at background and linear-perturbation level (equations of state, speeds
of sound, masses, coupling strengths, etc.) which then have some particular
effect on the phenomenology (evolution of the scale factor and perturbations
of matter and DE itself). In this paper, we employ a different approach,
aiming to answer the following questions: 
\begin{enumerate}
\item Assuming the ideal case of cosmological observations of the expansion
history and linear perturbations in a range of redshifts and scales
which are as precise as is required, which physical properties (such
as e.g.~the Hubble rate $H(z)$, the perturbation normalization $\sigma_{8}$,
the perturbation growth rate $f$, etc.) could actually be reconstructed
if we were to refrain from any parameterization of dark energy? 
\item Can we use these observable quantities to rule out not just some particular
cosmological model but the entire class of viable single scalar-field
models? 
\end{enumerate}
We are of course not the first to attempt to study dark energy cosmologies
in a model-independent way. In Ref.~\cite{Stebbins:2012vw} the author
argues for a  reconstruction method which does not rely even
on Einstein equations. This method can in principle directly measure
the space-time curvature but cannot test a modification of gravity.
On background level, the power of observations to constrain the expansion
history of the universe by assuming $\Lambda$-cold dark matter ($\Lambda$CDM)
as a null test was studied in Refs~\cite{Shafieloo:2009hi,Shafieloo:2011zv},
while a principle-component analysis of the equation of state constraints
was developed in Ref.~\cite{Huterer:2002hy}. Recently, even the
validity of the assumption of the Copernican principle was tested
\cite{Valkenburg:2012td}. On the level of linear perturbations, parameterizations
are usually used to limit the freedom in the model-independent description
of growth of structure \cite{Caldwell:2005ai,Linder:2005in,Hu:2007pj,Amendola:2007rr,Bean:2010zq},
although principle components analysis has also been employed \cite{Pogosian:2010tj}.
Another way to limit the freedom in the fully general description
is to exploit the structure which any general-relativity-like theory
of dark energy must obey: such approaches were discussed in \cite{Bertschinger:2006aw,Baker:2011jy,Battye:2012eu,Baker:2012zs}.
All of the above approaches, when contrasted with data, require parameterizations
in order to break degeneracies, but simultaneously introduce parameterization-dependent
biases.

The difference of our approach is that, given the minimum of assumptions,
we \emph{first} elucidate the observables that measurements can in
principle provide without the assumption of any dark energy model
in particular. It is only then that we use these model-independent
observables to construct tests which might eliminate or confirm particular
models. Our approach is closest in spirit to Ref.~\cite{Zhang:2007nk},
the results of which we extend.

In this paper, we completely ignore the practical problems and limitations
of the observations and assume that good-enough statistics with sufficiently
small systematic errors can be achieved in the range of redshifts
and scales discussed here. By exploring this idealized case we try
to discover the fundamental limits to which observations in a dark
energy cosmology are subject.

We adopt metric signature $(-+++)$, while a comma denotes a partial
derivative. We interchangeably use coordinate time $t$, scale factor
$a$, $e$-foldings $N\equiv\ln a$ and the redshift $z$ as time
variables. Overdots denote derivatives with respect to $t$, primes
with respect to $N$. The subscript $0$ denotes the present time.
We also make use of the notation of \cite{DeFelice:2011bh,DeFelice:2011hq},
to which we refer for a thorough study of linear perturbations in
the context of the Horndeski theories.

\section{Assumptions}

\label{sec:Assumpts}In keeping with the spirit of generality, we
first wish to make the minimum of assumptions on the geometry and
matter content of our Universe that will allow us to interpret observations
at all. In the following we assume only that: 
\begin{enumerate}[%
label=(\emph{\alph*})%
]
\item The geometry of the Universe is well described
by small linear perturbations living in an FLRW metric with scale
factor $a(t)$. We will not consider possible observations of rotational
perturbation modes nor of gravitational waves, as these are irrelevant
for structure formation in late-time cosmology. 
\item The matter content (i.e.~dark matter and baryonic matter) is pressureless
or evolves in a known way. 
\item The relation between the galaxy distribution and the matter distribution
at linear scales can be modeled as $\delta_{\text{gal}}=b(k,a)\delta_{\text{m}},\,$
where $b(k,a)$ is the potentially scale- and time-dependent linear
bias, while at the same time there is no bias between the velocities
of galaxies and matter. This implies that both the baryonic and dark
matter respond in the same way to the gravitational potentials and
that the statistical velocity bias due to galaxies sampling preferentially
over-dense regions \cite{Desjacques:2009kt} is negligible on the
scales of interest. 
\item The late-time universe is effectively described by the action 
\begin{equation}
S=\int\text{d}^{4}x\sqrt{-g}\left(\tfrac{1}{2}R+\mathcal{L}_{x}+\mathcal{L}_{\text{m}}\right)\,,\label{eq:Action}
\end{equation}
(setting $8\pi G_{\text{N}}=1$) which includes the Einstein-Hilbert
term for the metric $g_{\mu\nu}$ and the Lagrangian $\mathcal{\mathcal{L}_{\text{m}}}$
describing pressureless matter fluids, \emph{both} baryons and dark
matter, between which we will not differentiate here. Any other terms
are ascribed to the DE Lagrangian $\mathcal{L}_{x}$, which represents
some consistent theory potentially depending on extra degrees of freedom or $g_{\mu\nu}$ (i.e.~modifications of gravity).%
\footnote{A consistent theory is understood here to be a theory free of ghost
and other catastrophic instabilities that can in general occur in
generalized gravity and dark energy models.}
We will neglect the radiation component
because all the observations are assumed to be performed well after
decoupling. In non-minimally coupled models, the Lagrangian $\mathcal{L_{\text{m}}}$
depends on a different metric, related to $g_{\mu\nu}$ through some
transformation. Here we assume, however, that we have already reformulated
the action so that matter moves on the geodesics of $g_{\mu\nu}$. 
\end{enumerate}
We employ the above minimal framework to address question (1) by considering
the background observables in section \ref{sec:Bckg} and those arising
from linear perturbations in section \ref{sec:LinPerts}. To answer
question (2) we need another crucial assumption, concerning the degrees
of freedom in the dark energy Lagrangian: 
\begin{enumerate}[%
label=(\emph{\alph*}),start=5%
]
\item The Lagrangian $\mathcal{L}_{x}$, which
describes dark energy, is any one of the Lagrangians describing a
single scalar field governed by second-order equations of motion.
We call this scalar field dark energy, but we do not necessarily require
it to be driving the current acceleration. For example, it could be
that the late time acceleration is effectively driven by a cosmological
constant, but in the presence of this additional degree of freedom.
The assumption of a scalar field ensures that there are no gross violations
of isotropy. The limitation to second order is a necessary condition
to ensure that the model is not subject to instabilities (see e.g.~\cite{Woodard:2006nt}).%
\footnote{%
 This class of Lagrangians includes such theories as $f(R)$ gravity,
despite their naively fourth-order equations of motion. This is because
we can always introduce the a priori hidden scalar explicitly through
a Legendre transformation \cite{Chiba:2003ir}. %
} We will therefore assume that the dark
energy is governed by the most general Lagrangian which fulfills these
requirements: $\frac{1}{2}R+\mathcal{L}_{x}$ will form the Horndeski
Lagrangian (HL \cite{Horndeski:1974,Deffayet:2011gz}). We dedicate
section \ref{sec:Horn} to this system. 
\end{enumerate}

\section{Background Observables}

\label{sec:Bckg}From assumptions $(a)$-$(c)$, by varying
the action eq.~\eqref{eq:Action} with respect to the metric, we
obtain a Friedmann equation that can be written as 
\begin{equation}
H^{2}-H_{0}^{2}\Omega_{k0}a^{-2}=\frac{1}{3}(\rho_{x}+\rho_{\text{m}})\,,\label{eq:Friedmann}
\end{equation}
where $H_{0}$ is the present value of the Hubble parameter, $\Omega_{k0}$
the present curvature density parameter and $\rho_{\text{m}}$ is
the matter energy density. From assumption $(b)$, $\rho_{\text{m}}$
evolves as $a^{-3}$, and $\rho_{x}$ is the energy density of the
terms coming from $\mathcal{L}_{x}$.

Observations of the cosmic expansion are essentially estimations of
distances $D(z)$ (i.e.~luminosity or angular-diameter distances)
or directly $H(z)$ (e.g.\ using measurements of longitudinal baryon
acoustic oscillations, or real-time redshift-drift observations \cite{Quercellini:2010zr})
based on the existence of standard candles, rods or clocks. More exactly,
standard candles or rods measure $H(z)$ up to a multiplicative constant,
related to the unknown absolute measure of the source luminosity or
proper length. For instance, the flux of supernovae Ia (SNIa) with
absolute luminosity $L$ are known only up to the constant $LH_{0}^{2}$;
only ratios of fluxes at different redshifts are independent of the
absolute normalization. The same is true of baryon acoustic oscillations:
they measure only the ratio of the sound horizon at last scattering
and the Hubble radius $H^{-1}(z)$ . We can therefore say that, without
additional assumptions, background cosmological observations estimate
$D(z)$ up to an overall constant as well as the dimensionless Hubble
function $E(z)\equiv H(z)/H_{0}$. Notice, however, that real-time
redshift-drift observations can estimate the absolute value of $H(z)$,
while local measurements of the expansion measure $H_{0}$.

Combining $D(z)$ with $H(z)$, we can also estimate the present curvature
parameter $\Omega_{k0}$. We can therefore determine the evolution
of the combined matter and dark energy content, $1-\Omega_{k}$, at
all times. If we assume that there are only two components of the
cosmic fluid then we have only one free parameter, $\Omega_{\text{m}0}$.
In fact, we can write 
\begin{equation}
\Omega_{x}=1-\Omega_{k}-\Omega_{\text{m}}=1-\frac{1}{E^{2}}\left(\Omega_{k0}a^{-2}+\Omega_{\text{m}0}a^{-3}\right)\,.
\end{equation}
Therefore, we conclude that from background observables we can reconstruct
both $\Omega_{\text{m}}$ and $\Omega_{x}$, but only up to $\Omega_{\text{m}0}$
\cite{Kunz:2007rk}, since one can compensate for any change of $\Omega_{\text{m0}}$
with a modification of the DE model. Of course, if we parameterize
the evolution of $\Omega_{x}$ with a simple equation of state, we
can break the degeneracy with $\Omega_{\text{m}0}$, as is usually
done in analyses of SNIa data, but that is exactly what we are trying
to avoid in this work.

The same result is valid if instead of pure pressureless matter one
includes further components (e.g.~massive neutrinos) that evolve
with an effective equation of state $w_{\text{m}}(z)$, provided $w_{\text{m}}(z)$
can be inferred from other observations (e.g.~knowledge of the neutrino
masses).

\section{Linear Perturbation Observables}

\label{sec:LinPerts}The linear perturbation observables are
the correlations of positions, velocities and shapes (ellipticities)
of sources (i.e.~galaxies, Lyman-$\alpha$ lines, clusters, background
radiation) in angular separation and redshift. Given knowledge of
$D(z)$, these can be converted to the more usual dependence on wavenumber
$k$ and redshift.

Let us first discuss the clustering of matter. We denote the root
mean square of the correlation of galaxy number counts in Fourier
space as $\delta_{\text{gal}}$ (i.e.~$\delta_{\text{gal}}\equiv P_{\text{gal}}^{1/2}(k,z)$
where $P_{\text{gal}}$ is the galaxy power spectrum). We define from
now on the wavenumber $k$ to be the physical wavenumber expressed
in the units of the cosmological horizon, i.e.~$k=k_{\text{phys}}/aH$
($k$ is independent of $H_{0}$ if $k_{\text{phys}}$ is measured
in $h/$Mpc). This means that $k$ is time-dependent. We observe galaxies,
not matter perturbations, so as anticipated we need to introduce a
bias function $b(k,z)$ such that $\delta_{\text{gal}}=b\delta_{\text{m}}$.
Without the assumption of a particular model, DE perturbations are
unknown. In many models they are not at all small compared to matter
perturbations. For this reason, we will define the \emph{total density
}perturbation $\delta_{\text{t}}\equiv\Omega_{\text{m}}\delta_{\text{m}}+\Omega_{x}\delta_{x}$
and introduce the bias $B$ of galaxies with respect to it, 
\begin{equation}
\delta_{\text{gal}}=B\delta_{\text{t}}=BZ\Omega_{\text{m}}\delta_{\text{m}}\,,
\end{equation}
where $Z(k,a)\equiv1+\Omega_{x}\delta_{x}/(\Omega_{\text{m}}\delta_{\text{m}})$
is a function of space and time that depends on the clustering of
the $x$-component. Then we have that $b=BZ\Omega_{\text{m}}$.

Let us denote the initial total density perturbation spectrum at decoupling
as $\delta_{\text{t,in}}^{2}(k)$ and as $G_{\text{t}}(k,z)$ the
scale-dependent growth function of the linear \emph{total }density
perturbations, normalized to unity at present. If the galaxies move
with the same velocity field as matter, the galaxy velocity divergence
$\theta_{\text{gal}}$ in the sub-Hubble regime is related to the
\emph{matter }density perturbation as $\theta_{\text{gal}}=\theta_{\text{m}}=-\delta'_{\text{m}}=-f\delta_{\text{m}}$,
by the continuity equation for matter. We then obtain $\theta_{\text{gal}}=-(f/b)\delta_{\text{gal}}$,
where $f=G'/G$ is the linear matter growth rate and $G(k,z)$ is
the growth function for matter perturbations, both of which are scale-dependent
for general DE models. This velocity field generates redshift distortions
as a function of the direction cosine $\mu=(\vec{k}\cdot\vec{\ell})/k$
where $\vec{\ell}$ is the unit line-of-sight vector. The observable
$\delta_{\text{gal}}$ can therefore be expressed as \cite{Kaiser:1987qv}
\begin{equation}
\delta_{\text{gal}}(k,z,\mu)=G_{\text{t}}B\sigma_{8\text{,t}}\left(1+\frac{f}{b}\mu^{2}\right)\delta_{\text{t,0}}(k)\,,
\end{equation}
where $\sigma_{\text{8,t}}$ is the present normalization of the total
density spectrum. Now we can write $G_{\text{t}}B=Z\Omega_{\text{m}}\delta_{\text{m}}B/(Z\Omega_{\text{m}}\delta_{\text{m}})_{0}=Gb/(Z\Omega_{\text{m}})_{0}$,
so we have the almost-standard expression 
\begin{equation}
\delta_{\text{gal}}(k,z,\mu)=Gb\sigma_{8}\left(1+\frac{f}{b}\mu^{2}\right)\delta_{\text{t,}0}(k)\,.
\end{equation}
In this expression we set, using the 8 Mpc$/h$ spherical Fourier
space window function $W(k)$, 
\begin{eqnarray}
\sigma_{8}=\frac{\sigma_{8,\text{t}}}{(Z\Omega_{\text{m}})_{0}} & = & Z_{0}^{-1}\left(\int Z_{0}^{2}P_{\text{m}0}W^{2}(k,R_{8})\mathrm{d}k\right)^{1/2},
\end{eqnarray}
which is equivalent to the usual normalization $\sigma_{8}$ if $Z$
depends weakly on $k$.

To discuss weak lensing, we introduce the standard perturbed metric
in longitudinal gauge $\mathrm{d}s^{2}=-(1+2\Psi)\mathrm{d}t^{2}+a^{2}(1+2\Phi)\delta_{ij}\mathrm{d}x^{i}\mathrm{d}x^{j}$.
Since later we will confine ourselves to a scalar dark energy, as
part of assumption (\emph{a}), we assume that the DE does not excite
vector and tensor modes so that only scalar modes need to be included.
It is helpful to introduce the function $Y$, the effective gravitational
constant for matter, and the anisotropic stress $\eta$ defined as
(see e.g.\ \cite{Amendola:2007rr,DeFelice:2011hq}) 
\begin{align}
Y(k,a) & =-\frac{2k^{2}\Psi}{3\Omega_{\text{m}}\delta_{\text{m}}}\,,\quad\eta(k,a)=-\frac{\Phi}{\Psi}\,.\label{eq:poisson-1}
\end{align}
Both $\eta$ and $Y$ are unity on sub-Hubble scales if $\mathcal{L}_{x}$
is a constant (i.e.~the DE is a cosmological constant) and the matter
is a perfect fluid. In the linear regime, the lensing effect is proportional
to the lensing potential, which itself is driven by the density perturbations
(see e.g.~\cite{Dodelson2003}). In general, this relation can be
written as 
\begin{eqnarray}
k^{2}\Phi_{\text{lens}} & = & k^{2}(\Psi-\Phi)=-\frac{3}{2}Y(1+\eta)\Omega_{\text{m}}\delta_{\text{m}}\\
 & = & -\frac{3}{2}Y(1+\eta)\frac{G_{\text{t}}\sigma_{\text{8,t}}\delta_{\text{t,0}}}{Z}=-\frac{3}{2}\Sigma G\Omega_{\text{m}}\sigma_{8}\delta_{\text{t,0}}\,,\nonumber 
\end{eqnarray}
where we have defined the ``modified lensing'' function $\Sigma(k,z)\equiv Y(1+\eta)$.
The ellipticity correlation is an integral function of $\Phi_{\text{lens}}$
within a window function that depends on the survey geometry (see
e.g.~\cite{Zhang:2007nk}). Assuming good-enough knowledge of the
galaxy distribution one can differentiate the correlation integral
and obtain the quantity 
\begin{eqnarray}
\sigma(k,z) & \equiv & \frac{2}{3}(k^{4}P_{\Phi_{\text{lens}}})^{1/2}=\frac{1}{a^{3}E^{2}}\Omega_{\text{m}0}\Sigma G\sigma_{8}\delta_{\text{t,0}}\,.
\end{eqnarray}
Then from $\delta_{\text{gal}}(k,z,\mu)$ (with e.g.~$\mu=0,1$)
and $\sigma(k,z)$ one can measure the three quantities $A,R,L$ defined
as 
\begin{align}
A & =Gb\sigma_{8}\delta_{\text{t,0}}\,,\quad R=Gf\sigma_{8}\delta_{\text{t,0}}\,,\\
L & =\Omega_{\text{m0}}\Sigma G\sigma_{8}\delta_{\text{t,0}}\,.
\end{align}
The quantities that connect the observations to theory (i.e.~to the
Lagrangian $\mathcal{L}_{x}$) are $\Omega_{x}$, $f$, $\Sigma$,
so it would be optimal to estimate them directly from observations.
Now, the cosmic microwave background anisotropy allows one to measure,
at least in principle, the initial potential $\Psi_{\text{in}}$ through
the Sachs-Wolfe effect. It is, however, impossible to derive from
this information the present power spectrum $\delta_{\text{t,0}}$
since it also depends on a scale- and time-dependent transfer function.
Absent a model for DE, this transfer function is unknown and since
it acts to process the total perturbation spectrum, changing its $k$-dependence,
it also makes $\delta_{\text{t,0}}$ an unknown without further assumptions.
This argument shows that the only $\delta_{\text{t,0}}$-independent
quantities directly measurable from linear cosmological observations
are ratios of $A,R,L$, and their $N$-derivatives, i.e. 
\begin{eqnarray}
P_{1} & \equiv & R/A=f/b,\\
P_{2} & \equiv & L/R=\Omega_{\text{m}0}\Sigma/f,\\
P_{3} & \equiv & R'/R=f+f'/f.
\end{eqnarray}
All other possible $\delta_{\text{t,0}}$-independent ratios, such
as $A'/A$, $L'/L$ or $R'/L$ or higher-order $N$-derivatives, can
be obtained as combinations of $P_{1-3}$ and their derivatives, for
instance $L'/L=P_{2}'/P_{2}+P_{3}$. Other linear-perturbation probes,
such as integrated Sachs-Wolfe, cross-correlations or 21-cm flux measurements,
add statistics and might extend the observational range but do not
break the fundamental degeneracy.

The quantity $P_{1}$ is a well-known observable quantity, often denoted
$\beta$ \cite{Dodelson2003}; since it involves the bias function
$b$, related in an unknown way to the model of dark energy, we will not consider it any longer in this paper. The quantity $P_{2}$
has already been introduced in \cite{Zhang:2007nk} as $E_{G}$ as
a test of modified gravity, but the fact that $\Omega_{\text{m0}}$
is not an observable was not discussed there. The quantity
$R$ contains the term $Gf\sigma_{8}$, also denoted as $f\sigma_{8}(z)$
in the literature \cite{Percival:2008sh}. This term is often considered
to be a directly observable quantity, but, as we have argued, this
is only true if one assumes a model for DE, or at least a parametrized
form of $\delta_{\text{t},0}$; otherwise, the model-independent observable
combination is $P_{3}=R'/R$. It is important to realize that even
a perfect knowledge of $P_{3}$ does not imply knowledge of $f$ since
the equation $f'/f+f=P_{3}(k,z)$ cannot be solved without the unknown
$k$-dependent initial condition for $f$. Finally, notice that we
did not need to assume Gaussian fluctuations nor isotropy of the power
spectrum.

Measurements of galaxy peculiar velocities and their time derivative
directly estimate $\Psi$ through the Euler equation, which would
give the quantity $V=\Omega_{\text{m0}}YG\sigma_{8}\delta_{\text{t,0}}\,$.
Then one can form the observable, $L/V=1+\eta$, which measures the
anisotropic stress. This new observable is not independent since it
can be written in terms of $E,P_{2},P_{3}$ (see Eq.~\eqref{eq:kpol}).
Moreover, the estimation of $V$ requires a delicate subtraction of
the peculiar redshift from the cosmological redshift by using distance
indicators such as Cepheids and therefore a number of additional assumptions
on the source physics. No current or foreseeable method to estimate
the peculiar velocity field (let alone its derivative) has been shown
to be reliable beyond a few hundred megaparsecs (see e.g.~Ref.~\cite{Macaulay:2011av}),
so we will not pursue this possibility any further in this paper.

Our first result is that linear cosmological observations can at best
determine only $E\equiv H/H_{0}$ as function of time (as well as
$\Omega_{k0}$ but \emph{not} $\Omega_{\text{m}0}$) and the observable combinations $P_{1-3}$ as functions
of time and space, within the range of the observations themselves.
To achieve this, we need to combine galaxy clustering and weak lensing
in the same redshift range. It is possible that one can determine
other combinations from non-linear effects, but this will certainly
introduce new uncertainties (e.g.~a non-linear bias).

\section{The Horndeski Lagrangian}

\label{sec:Horn}It is now time to use also assumption $(e)$
regarding the explicit form of the dark energy Lagrangian. In the
choice of the action \eqref{eq:Action}, we have assumed that all
matter components (i.e.~dark matter and baryons) feel the same gravitational
force and propagate on geodesics of the metric $g_{\mu\nu}$. In addition,
we now explicitly assume that the DE is modeled by a single scalar
field $\phi$ described by the Horndeski Lagrangian (HL). The HL is
defined as the sum of four terms $\mathcal{L}_{2}$ to $\mathcal{L}_{5}$
that are fully specified by a non-canonical kinetic term $K(\phi,X)$
and three in principle arbitrary coupling functions $G_{3,4,5}(\phi,X)$,
where $X=-g_{\mu\nu}\phi^{,\mu}\phi^{,\nu}/2$ is the canonical kinetic
term,

\begin{align}
\begin{aligned}\mathcal{L}_{2}= & K(\phi,X)\,,\\
\mathcal{L}_{3}= & -G_{3}(\phi,X)\Box\phi\,,\\
\mathcal{L}_{4}= & G_{4}(\phi,X)R+G_{4,X}\left[\left(\Box\phi\right)^{2}-\left(\nabla_{\mu}\nabla_{\nu}\phi\right)^{2}\right]\,,\\
\mathcal{L}_{5}= & G_{5}(\phi,X)G_{\mu\nu}\nabla^{\mu}\nabla^{\nu}\phi-\frac{G_{5,X}}{6}\Bigl[\left(\Box\phi\right)^{3}-\\
 & -3\left(\Box\phi\right)\left(\nabla_{\mu}\nabla_{\nu}\phi\right)^{2}+2\left(\nabla_{\mu}\nabla_{\nu}\phi\right)^{3}\Bigr]\,.
\end{aligned}
\label{eq:HL}
\end{align}
The Horndeski Lagrangian is the most general Lagrangian for a single
scalar which gives second-order equations of motion for both the scalar
and the metric on an arbitrary background. This is a necessary, but
not sufficient, condition for the absence of ghosts.%
\footnote{The constraints on the HL arising from stability considerations are
derived in Ref.~\cite{DeFelice:2011bh}. We have also presented them
in appendix \ref{sec:App}.%
} In general, the equation of motion for the scalar will couple it to
the matter energy density. The metric potentials $\Phi$ and $\Psi$
are as usual determined by the Poisson and anisotropy equations, which
are constraints, and therefore do not have independent dynamics. We
note that the generalization to the case of multiple scalar fields
has been discussed in Ref.~\cite{Padilla:2012dx}.

In what follows, we will assume that the so-called \emph{quasi-static
limit} is valid for the evolution of perturbations. This implies that
we are observing scales significantly inside the cosmological horizon,
$k\equiv k_{\text{phys}}/(aH)\gg1$, and inside the Jeans length of
the scalar, $c_{\text{s}}k\gg1$, such that the terms containing $k$
dominate over the time-derivative terms. The sound speed $c_{\text{s}}$
is a particular function of the HL functions $K,G_{3-5}$ evaluated
at the \emph{background }level and we have presented it in Eq.~\eqref{eq:c_s} \cite{DeFelice:2011bh}. In this quasi-static limit for a model belonging
to the HL, one obtains \cite{DeFelice:2011hq} 
\begin{align}
\eta & =h_{2}\left(\frac{1+k^{2}h_{4}}{1+k^{2}h_{5}}\right)\,,\, Y=h_{1}\left(\frac{1+k^{2}h_{5}}{1+k^{2}h_{3}}\right)\,.\label{eq:etay}
\end{align}
for suitably defined functions $h_{1-5}.$ In this limit, one also
has $Z=Y\eta$.

The functions $h_{1-5}$ express the modification of gravity induced
by the HL. In real space they induce a time-dependent Yukawa correction
to the Newtonian potential. They are all combinations of HL functions
$K,G_{3,4,5}$ and their derivatives with respect to $\phi$ and $X$,
all evaluated on the background and are therefore time- but not $k$-dependent,
\[
h_{i}\equiv h_{i}(z)\equiv h_{i}(\phi,X).
\]
The explicit expressions for the functions $h_{i}$ are very complicated
and not particularly illuminating; we have nonetheless presented them
in appendix \ref{sec:App}. For $\Lambda$CDM one has simply $h_{1,2}=1$
and $h_{3,4,5}=0$. If the two gravity-coupling functions in the HL,
$G_{4},G_{5}$, are constant, i.e.~the effective Planck mass is constant,
then $\eta=1$; if moreover $G_{3}$ depends only on $\phi$ (i.e.~k-\emph{essence}),
then also $Y=1$ and there are no modified-gravity effects at all
in this quasi-static limit.

It is worth noting that one could have arrived at the form of Eqs~\eqref{eq:etay}
given just our assumptions of second-order equations of motion, the
symmetries of the FRW background and quasi-staticity.

In the same quasi-static limit, from the matter conservation equation,
we obtain 
\begin{equation}
\delta_{\text{m}}''+\left(2+\frac{H'}{H}\right)\delta_{\text{m}}'=-k^{2}\Psi=\frac{3}{2}\Omega_{\text{m}}\delta_{\text{m}}h_{1}\left(\frac{1+k^{2}h_{5}}{1+k^{2}h_{3}}\right)\,,
\end{equation}
or 
\begin{align}
f'+f^{2}+\left(2+\frac{H'}{H}\right)f & =\frac{3}{2}\Omega_{\text{m}}h_{1}\left(\frac{1+k^{2}h_{5}}{1+k^{2}h_{3}}\right)\,.\label{eq:f}
\end{align}
On the other hand, we can write for the weak-lensing function $\Sigma$,
\begin{equation}
\Sigma=Y(1+\eta)=h_{6}\left(\frac{1+k^{2}h_{7}}{1+k^{2}h_{3}}\right)\,,\label{eq:sigma}
\end{equation}
where we have introduced two auxiliary functions, $h_{6}=h_{1}(1+h_{2})$
and $h_{7}=(h_{5}+h_{4}h_{2})/(1+h_{2})$. As an aside, one can show
that if $G_{3},G_{4}$ depend only on $\phi$ and $G_{5}=\text{const}$,
i.e.~we are dealing with a k-\emph{essence} theory non-minimally
coupled to gravity, then $h_{2}=1$ and $h_{7}=h_{3}$ so that $\Sigma$
becomes independent of $k$. In this limit, the gravitational potential
felt by photons is not distorted (we have discussed this model in
detail in Ref.~\cite{Sawicki:2012re}). 

From the observables $P_i$, $E$ we can construct a model-independent relation measuring the anisotropic stress $\eta$ as follows. From $P_{2},P_{3}$ we can obtain $f=\Omega_{\text{m}0}\Sigma/P_{2}$
and $f'=P_{3}\Omega_{\text{m}0}\Sigma/P_{2}-(\Omega_{\text{m}0}\Sigma/P_{2})^{2}$.
Inserting this in Eq.~\eqref{eq:f} and employing Eq.~\eqref{eq:sigma},
after a little algebra we obtain a simple relation.
\begin{equation}
\frac{3P_{2}(1+z)^{3}}{2E^{2}\left(P_{3}+2+\frac{E'}{E}\right)}-1 =\eta =h_{2}\left(\frac{1+k^{2}h_{4}}{1+k^{2}h_{5}}\right)\,.\label{eq:kpol}
\end{equation}
It is important to stress that the l.h.s.~of Eq.~\eqref{eq:kpol} is a 
function of model-independent observables, and thus a model-independent measurement 
of $\eta$, valid under our assumptions (\emph{a})-(\emph{d}), but \emph{not} requiring (\emph{e}). 
The form of the last term is determined by the QS limit of the Horndeski Lagrangian. Given the above, 
we can exclude all dark energy models described by a single 
scalar field in the QS limit by showing that the anisotropic stress measured 
from the observation data does not follow the particular scale dependence mandated by Eq.~\eqref{eq:kpol}.
Equation~\eqref{eq:kpol} must be valid in fact at all times and scales where the quasi-static
limit is valid. At any given epoch $z^{*}$, this equation involves
the three unknowns $h_{2},h_{4,}h_{5}$ all evaluated at $z^{*}$.
If at this epoch we observe $E$ and $P_{2},P_{3}$ at more than three
$k$-modes, we can form an overconstrained system. If for any $z^{*}$
this system has no solution then the observations are inconsistent
with the quasi-static limit of HL. Equivalently, from Eq.~\eqref{eq:kpol}
one can obtain a consistency relation that depends only on observable
quantities. Defining $g(z,k)\equiv\frac{(REa^{2})'}{LEa^{2}}$, one
has in fact 
\begin{equation}
2g^{(1)}g^{(3)}-3(g^{(2)})^{2}=0\,,\label{eq:cons}
\end{equation}
where $g^{(n)}$ is the $n$-th derivative of $g$ with respect to
$k^{2}$. If this condition fails at any one redshift, the
DE is not described by the HL in the linear quasi-static limit. This
is the second main result of this paper. Needless to say, a cosmological
constant satisfies this consistency relation.

On the other hand, if there are consistent solutions then we obtain
an indication in favor of the HL and also direct constraints on it.
For instance, if observationally we find that $P_{2},P_{3}$ do not
depend on $k$, then from Eq.~\eqref{eq:kpol} we see that the condition
$h_{4}=h_{5}$ must be satisfied.

If the consistency relation is not satisfied, the only possible way
out of our conclusion is that the conditions for the linear quasi-static
limit that we employed to derive Eq.~\eqref{eq:etay} are not satisfied.
This can occur if the rate of change of the functions $h_{1-5}$ is
very large, e.g.~if $h_{j}'/h_{j}\approx c_{\text{s}}^{2}k^{2}\gg1$
for some $j$. However if the field $\phi$ drives the current accelerated
expansion (this is indeed an additional assumption) we expect it to
be slow rolling on time scales of the order of $\dot{h}_{j}/h_{j}\sim H$,
i.e.~$h_{j}'/h_{j}\approx1$. If the sound speed squared $c_{\text{s}}^{2}$
is of order unity, then on typical astrophysical scales of $100$~Mpc$/h$
one has $k^{2}\approx10^{3}$, so the quasi-static limit should be
very well satisfied. However if $c_{\text{s}}^{2}$ is less than say
$10^{-2}$ then the simple form of Eq.~\eqref{eq:etay} is no longer
valid. One might then expect oscillating terms in the $Y,\eta$ equations;
it is possible that this behavior could be probed, and possibly rejected,
by a similar method we are discussing here but a full analysis of
this ``cold dark energy'' scenario would be required. Another potential
difficulty is the fact that these more general scalar field theories
contain non-linearities in principle independent of those in the matter
perturbations. It may prove difficult to determine on which scales
the linear approximation for the dark energy is valid, if at all.

From $P_{2}$ and $P_{3}$ one can build other consistency equations,
e.g.~by differentiating $P_{2}$ or the combination $P_{2}'/P_{2}+P_{3}=\Sigma'/\Sigma+f$
with respect to $N$ and again employing Eqs~\eqref{eq:f},~\eqref{eq:sigma}.
These relations however require derivatives of the observables $P_{2},P_{3}$
and will introduce derivatives of the $h_{1-5}$ functions, so appear
to be less useful than Eq.~\eqref{eq:kpol}.

We observe also that the propagation speed $c_{\text{T}}$ of gravitational
waves is a function of the HL coupling functions, see Eq.~\eqref{eq:c_T}
\cite{Kobayashi:2011nu}. A detection of a source both in gravitational
and electromagnetic waves could allow for a measurement of $c_{\text{T}}$
and therefore new independent constraints on the HL \cite{Kimura:2011qn}.
\\
\section{Conclusions}

We have shown that cosmological linear observations can measure only
$\Omega_{k0}$, $E\equiv H/H_{0}$ and the combinations $P_{1}=f/b$,
$P_{2}=\Omega_{\text{m}0}\Sigma/f$ and $P_{3}=f+f'/f$. Parameters
such as $\Omega_{\text{m}0}$, $\sigma_{8}$ or functions such as
$Gf\sigma_{8}$ are not directly model-independent measurable quantities
via linear cosmological observations alone. This limits in a fundamental
way the knowledge of, among others, the evolution of the DE density
parameter $\Omega_{x}$, its equation of state, or the matter growth
rate $f$. From $E,P_{2,3}$ one can form consistency relations in
terms of the HL functions. The simplest one is Eq.~\eqref{eq:kpol}
or \eqref{eq:cons}, expressed purely in terms of model-independent
observables. If observations indicate a violation of a consistency
relation, then the DE is not described by the HL in the quasi-static
limit. Conversely, finding the predicted $k$-behavior would be a
major confirmation of the scalar field picture of dark energy.

Non-linear effects will bring both more information and more unknowns
into the picture so it is not clear how much, if at all, they would
improve the task of reconstructing or rejecting the HL. The limitations
of real-world observations, completely neglected here, are of course
in practice the major hurdle on the path to this goal. 
\begin{acknowledgments}
It is a pleasure to thank Leonidas Christodoulou, Enrique Gaztañaga,
Dragan Huterer, Eyal Kazin, Ofer Lahav and Shinji Tsujikawa for useful
comments and conversations. We are grateful for the hospitality of
the Centro de Sciencias de Benasque Pedro Pascual, where a part of
this manuscript was prepared. The work of L.A.~and I.S.~is supported
by the DFG through TRR33 ``The Dark Universe''. M.K.~acknowledges
funding by the Swiss National Science Foundation. M.M.~is supported
by CNPq-Brazil. I.D.S.~acknowledges STFC for financial support. 
\end{acknowledgments}
\appendix
\begin{widetext}
\section{Details of Horndeski Properties}

\label{sec:App}This appendix concerns the properties of the
scalar-field theories described by the Horndeski Lagrangian given
by the combination of terms presented in Eqs~\eqref{eq:HL}. The
subscripts $,\phi$ and $,X$ denote derivation with respect to that
variable. On a flat FRW background, the energy density and pressure
are given by

\begin{align}
\rho_{x}= & 3H^{2}(1-w_{1})+2XK_{,X}-K-2XG_{3,\phi}+\\
 & +6\dot{\phi}H\left(XG_{3,X}-G_{4,\phi}-2XG_{4,\phi X}\right)+\nonumber \\
 & +12H^{2}\left(X\left(G_{4,X}+2XG_{4,XX}\right)-G_{5,\phi}-XG_{5,\phi X}\right)+\nonumber \\
 & +4\dot{\phi}XH^{3}\left(G_{5,X}+XG_{5,XX}\right)\,,\nonumber \\
P_{x}= & -\left(3H^{2}+2\dot{H}\right)(1-w_{1})+K-2XG_{3,\phi}+4XG_{4,\phi\phi}+\nonumber \\
 & +2\dot{\phi}Hw_{1,\phi}-4X^{2}H^{2}G_{5,\phi X}+2\dot{\phi}XH^{3}G_{5,X}+\frac{\ddot{\phi}}{\dot{\phi}}\left(w_{2}-2Hw_{1}\right)\,,\nonumber 
\end{align}
where, given a slight rearrangement of the results in Refs~\cite{DeFelice:2011bh,DeFelice:2011hq}, we define four functions $w_i$ as

\begin{align}
w_{1}\equiv & 1+2\left(G_{4}-2XG_{4,X}+XG_{5,\phi}-\dot{\phi}XHG_{5,X}\right)\,,\label{eq:ws}\\
w_{2}\equiv & -2\dot{\phi}\left(XG_{3,X}-G_{4,\phi}-2XG_{4,\phi X}\right)+\nonumber \\
 & +2H\left(w_{1}-4X\left(G_{4,X}+2XG_{4,XX}-G_{5,\phi}-XG_{5,\phi X}\right)\right)-\nonumber \\
 & -2\dot{\phi}XH^{2}\left(3G_{5,X}+2XG_{5,XX}\right)\,,\nonumber \\
w_{3}\equiv & 3X\left(K_{,X}+2XK_{,XX}-2G_{3,\phi}-2XG_{3,\phi X}\right)+18\dot{\phi}XH\left(2G_{3,X}+XG_{3,XX}\right)-\nonumber \\
 & -18\dot{\phi}H\left(G_{4,\phi}+5XG_{4,\phi X}+2X^{2}G_{4,\phi XX}\right)-\nonumber \\
 & -18H^{2}\left(1+G_{4}-7XG_{4,X}-16X^{2}G_{4,XX}-4X^{3}G_{4,XXX}\right)-\nonumber \\
 & -18XH^{2}\Bigl(6G_{5,\phi}+9XG_{5,\phi X}+2X^{2}G_{5,\phi XX}\Bigr)+\nonumber \\
 & +6\dot{\phi}XH^{3}\Bigl(15G_{5,X}+13XG_{5,XX}+2X^{2}G_{5,XXX}\Bigr)\,,\nonumber \\
w_{4}\equiv & 1+2\left(G_{4}-XG_{5,\phi}-XG_{5,X}\ddot{\phi}\right)\,.\nonumber 
\end{align}
All of the dynamics of linear perturbations are fully determined by
the above four functions. In particular, the speed of propagation
of gravitational waves, $c_{\text{T}}$, and the normalization of
the kinetic term of these tensor perturbations, $Q_{\text{T}}$, is
given by

\begin{equation}
c_{\text{T}}^{2}=\frac{w_{4}}{w_{1}}>0,\qquad Q_{\text{T}}=\frac{w_{1}}{4}>0\,,\label{eq:c_T}
\end{equation}
with positivity required by stability. From the above, is can be seen
that $w_{1}$ has the meaning of the normalization of the tensor perturbations, i.e.~it is the effective Planck mass squared.
The corresponding quantities for the scalar degree of freedom, the
sound speed of dark energy, $c_{\text{s}}$, and the normalization
of the kinetic energy for perturbations, $Q_{\text{S}}$, in the presence
of dust with energy density $\rho_{\text{m}}$, are

\begin{align}
c_{\text{s}}^{2}= & \frac{3\left(2w_{1}^{2}w_{2}H-w_{2}^{2}w_{4}+4w_{1}w_{2}\dot{w_{1}}-2w_{1}^{2}(\dot{w}_{2}+\rho_{\text{m}})\right)}{w_{1}(4w_{1}w_{3}+9w_{2}^{2})}>0\,,\label{eq:c_s}\\
Q_{\text{S}}= & \frac{w_{1}\left(4w_{1}w_{3}+9w_{2}^{2}\right)}{3w_{2}^{2}}>0\,.\nonumber 
\end{align}
With above definitions in hand, we can define the five scale-independent
functions $h_{1-5}$ which appeared in the result \eqref{eq:etay}.
All the observables for scalar perturbations in the \emph{quasi-static} regime
are determined by these five functions,

\begin{align}
h_{1} & \equiv\frac{w_{4}}{w_{1}^{2}}=\frac{c_{\text{T}}^{2}}{w_{1}}\,,\qquad h_{2}\equiv\frac{w_{1}}{w_{4}}=c_{\text{T}}^{-2}\,,\label{eq:h1h2}\\
h_{3} & \equiv\frac{H^{2}}{2XM^{2}}\frac{2w_{1}^{2}w_{2}H-w_{2}^{2}w_{4}+4w_{1}w_{2}\dot{w}_{1}-2w_{1}^{2}(\dot{w}_{2}+\rho_{\text{m}})}{2w_{1}^{2}}\,,\nonumber \\
h_{4} & \equiv\frac{H^{2}}{2XM^{2}}\frac{2w_{1}^{2}H^{2}-w_{2}w_{4}H+2w_{1}\dot{w}_{1}H+w_{2}\dot{w_{1}}-w_{1}(\dot{w}_{2}+\rho_{\text{m}})}{w_{1}}\,,\nonumber \\
h_{5} & \equiv\frac{H^{2}}{2XM^{2}}\frac{2w_{1}^{2}H^{2}-w_{2}w_{4}H+4w_{1}\dot{w}_{1}H+2\dot{w_{1}}^{2}-w_{4}(\dot{w}_{2}+\rho_{\text{m}})}{w_{4}}\,,\nonumber 
\end{align}
and where the effective mass squared, $M^{2}$, can be expressed in
terms of derivatives of the total pressure and total energy with respect
to the scalar as 
\begin{equation}
M^{2}=\frac{3H\left(P_{x,\phi}+\rho_{x,\phi}\right)+\dot{\rho}_{x,\phi}}{\dot{\phi}}\,.
\end{equation}
One may wonder whether it is possible to invert Eqs~\eqref{eq:h1h2}
in order to obtain $\rho_{\text{m}}$ as a function of the functions
$h_{1-5}$. If this were possible, and if all the $h_{1-5}$ were
observable, then one could measure $\Omega_{\text{m}0}$, contrary
to our claim in the text. In a future paper we will discuss in detail
the observability of the functions $h_{1-5}$ and we will show that
in fact it is not possible to obtain $\Omega_{\text{m}0}$ this way.
Here it will suffice to notice that $h_{1}$ is completely degenerate
with $\Omega_{\text{m}0}$ (see Eq.~\eqref{eq:f}); it turns out
that $\rho_{\text{m}}$ is proportional to $1/h_{1}$ and therefore
the fact that $h_{1}$ can only be measured up to $\Omega_{\text{m}0}$
implies the same degeneracy in $\rho_{\text{m}}$. 
\end{widetext}
\bibliographystyle{utcaps}
\bibliography{observables}

\end{document}